# Error Control Codes: A Novel Solution for Secret Key Generation and Key Refreshment Problem

Arjun Puri
Assistant Professor, Baba Ghulam Shah Badshah University, Rajouri ,
Jammu and Kashmir.

Sudesh Kumar
Assistant Professor, Shri Mata Vaishno Devi University, Katra, Jammu and Kashmir

## ABSTRACT
Cryptography is the science of encrypting the information so that it is rendered unreadable for an intruder. Cryptographic techniques are of utmost importance in today's world as the information to be sent might be of invaluable importance to both the sender and the receiver. Various cryptographic techniques ensure that even if an intruder intercepts the sent information, he is not able to decipher it thus render ending it useless for the intruder. Cryptography can be grouped into two types, that is Symmetric key cryptography and Asymmetric key cryptography. Symmetric key cryptography uses the same key for encryption as well as decryption thus making it faster compared to Asymmetric Key cryptography which uses different keys for encryption and decryption. Generation of dynamic keys for Symmetric key cryptography is an interesting field and in this we have tapped this field so as to generate dynamic keys for symmetric key cryptography. In this work, we have devised an algorithm for generating dynamic keys for sending messages over a communication channel and also solving key refreshment problem.

## 1. INTRODUCTION
In communication, one of the main aspects is to send message correctly from sender to receiver. In order to achieve this, coding theory comes into picture that takes the responsibility of sending data correctly over noisy channel. There are mainly two types of coding theory one is forward error correction and another is backward error correction or automatic repeat request (ARQ). Another main duty of data communication with the advent of time is to provide security of the message to be sent. In order to send message securely, cryptography is most widely used. Cryptography is derived from two Greek words "*crypto*" which means secret and "*graphy*" which means writing. So this is known as secret writing. Cryptography is now an emerging research area where the scientists are trying to develop some good algorithms, for encryption and decryption which are hard to break by third body. Symmetric cryptography is also referred to as single key or shared key cryptography because only one key is used between sender and receiver in order to encrypt and decrypt the message. Symmetric key was developed in 1970. In Symmetric key cryptography plaintext can be encrypted by two ways: stream cipher and block cipher [1]. There are two major task involved in symmetric cryptography one is to generate key and another is to distribute key. In this paper we proposed an algorithm to generate key with the help of forward error correction codes. In order to achieve this, we have used Reed Solomon codes for the generation of key which in turn is used to encrypt message in cipher form. This paper is divided into following subsections:-related work, proposed work, implementation, conclusion and future work.

## 2. RELATED WORK
In [1], authors deal with a new symmetric key cryptographic method using dynamic key. In their work, they have used the Linear Congruential Generator (LCG) for generating key. They suggested the use of Linear Congruential Generator (LCG) for generating key. The advantage of the method is that every time new key is generated. So the process is very hard to break. The proposed algorithm is also compared with the other algorithms and it is concluded that performance of proposed algorithm is higher than other algorithms.

In [3], authors gave a novel approach to generate key based on face features. Traditional cryptographic methods require the user to remember keys. Thus it requires lot of storage space. So the concept of generating biometric key emerge which is much more secure as compared to traditional cryptographic method. In this paper, authors proposed biometric cryptosystem based on face biometric. At this phase, 128-dimensional principal component analysis (PCA) feature vector is firstly extracted from the face image and a 128 bit binary vector is obtained by thresholding. After that authors select the distinguishable bits to form bio-key and the optimal bit order number is saved in a look-up table. Finally they provide error correction by applying of Reed Solomon algorithm. The message is encrypted by using DES with bio-key. The proposed algorithm shows that the entropy of this algorithm is maximum and performance of this algorithm is better than other algorithm.

In [2] author suggests a new approach to protect the message by the application of symmetric cryptographic check values as message authentication code, over noise channel. In this method author introduce concept of hard verification of message authentication code.

In [4], author suggested that it is impossible to store all keys in case of large network. A natural solution then is to supply each user with a relatively small amount of secret data from which he can derive all his keys. In this paper he main emphases on the use of MDS codes.

In [5], authors explore a unique approach to generation of key using fingerprint. The generated key is used as an input key to the DES Algorithm. . The algorithm is successfully implemented and tested in Matlab. The paper also describes some of the commonly used technique for generation and extraction of minutiae points from fingerprint.

In [6], an implementation of the technique Matrix Embedding using the Reed-Solomon codes is obtained. The advantage of these codes is that they allow easy way to solve the problem of bounded syndrome decoding, a problem which is the basis of the technique of embedding matrix. The use of the technique of matrix embedding in the steganography makes it possible to minimize the changes introduced into the image of cover.

In [7], presents the implementation of Elliptic Curve Cryptography by first transforming the message (text and image) into an affine point on the Elliptic Curve over the finite prime field. The process of encryption of a text and image message and then a method for enhancing security using Forward Error Correction code for better error checking at the receiver side. This enables errors or noise added during transmission to be detected and corrected so that the receiver is receiving correct data and the received data matches with the sent data.

## 3. PROPOSED WORK

In the present scenario of cryptography, one of the important issues is to generate dynamic key because it is not possible all the time to store large amount of keys in order to secure communication. So it is important for us to generate keys dynamically. Key generation usually depends upon two major aspects:-Key stability and Key entropy. Key stability refers to the extent to which the key generated from the algorithm is repeatable. Key entropy refers to the numbers of dynamically generated keys.

These two factors are inversely propositional to each other. As we go on increasing Key stability, key Entropy goes on decreasing and vice-versa. In our present work we propose, "A Symmetric key generation model using forward error correction codes". In our proposed work we are generating keys with the help of one of the forward error codes. The proposed algorithm is given below:-

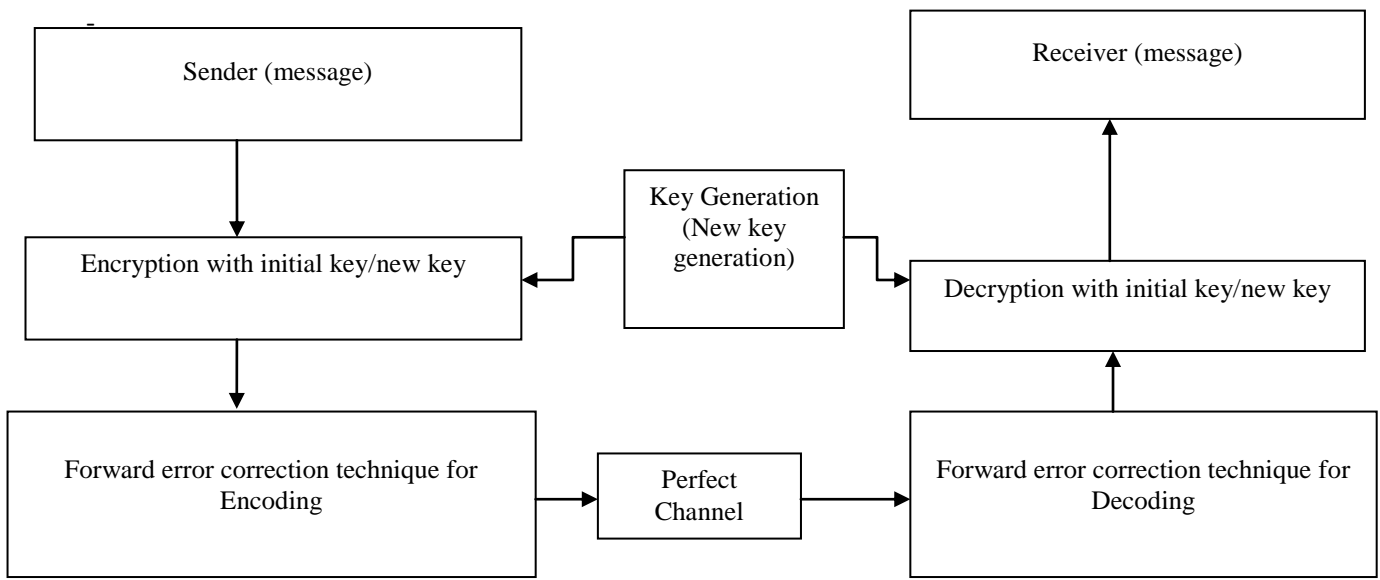

**Figure 1: - Proposed Model**

In first block, it takes plain text variable length from the user and converts it into ASCII code value and transfers to encryption. In encryption block, it is divided into small chunks and encryption is done along with key generation methods. After encryption, message will be converted into cipher text form. This whole process is performed at sender side. After that decryption is performed at receiving end. In decryption block, cipher text is converted into plaintext. At plain text of variable length, receives message in a confidential way.

To begin with, in key generation block we are trying to generate a new key for encryption. This is done by adding the initial key with message parity check matrix of forward error correction. The block diagram of the process is shown below:-

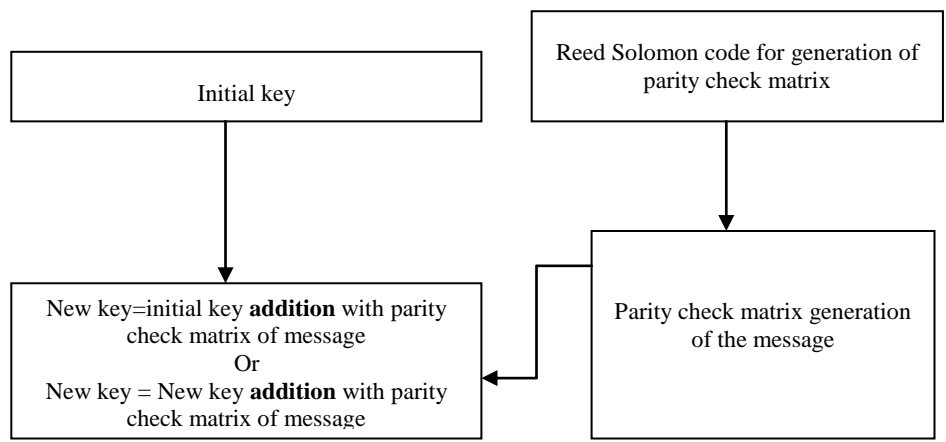

**Figure 2:-Key Generation Block**

## 4.      PROPOSED ALGORITHM

**Algorithm for key generation at sender side:-**
Step1: Check (length (input message)>63) then

Step2: Break it into message_chunks of 63 bits.

Then take individual chunks.

Step3: Take first chunks and perform encryption

Step4: [Encrypt] = input message + initial key

Step5: Perform RS encoding.

Step6: Extract Parity check matrix of first encoded message.

Step7: New key= initial key + parity check matrix of first message (used for encryption of second message and also perform RS encoding).

Step8: New key= new key + parity check matrix of second message (used for encryption of third message and also perform RS encoding).

And so on.

**Algorithm for key generation at receiver side:-**
Step1: Receive individual encoded chunks.

Step2: S= separate parity check matrix and store it for further decryption.

Step3: Take encrypted chunks from encoded chunk.

Step4: Then perform first decryption by using initial key.

Step5: After take further decryption of chunks by using algorithm

Step6:  Msg= encrypt- initial key

Step7: For further Msg=encrypt- new key

Where new key= initial key+ parity check matrix of first chunk

           OR

New key=new key+ parity check matrix

And so on.

## 5.      IMPLEMENTATION

The proposed work is implemented in Matlab. In implementation we have taken a message of length 63 characters and Reed Solomon code of size (127, 63). Input message having length greater than 63 is inputted to the system and its corresponding key which is used for encryption and decryption is generated at both ends as shown in figure 2 above. Firstly input message is converted into their respective ASCII code value then first chunk encryption is performed by using initial key after that further encryption is performed by generating keys from parity check matrix of the message and initial key. One example is shown as below:

**Input message**
In our simulation we took the following message "*rajasthan university and shri mata vaishno devi university are two good university in india*". The intermediate calculations carried out by the system is shown below

**ASCII code**
a =

  Columns 1 through 27

114    97   106    97   115   116   104    97   110    32   117
110   105   118   101   114   115   105   116   121    32    97
110   100    32   115   104

  Columns 28 through 54

114   105    32   109    97   116    97    32   118    97   105   115
104   110   111    32   100   101   118   105    32   117   110
105   118   101   114

  Columns 55 through 81

115   105   116   121    32    97   114   101    32   116   119
111    32   103   111   111   100    32   117   110   105   118
101   114   115   105   116

  Columns 82 through 92

121    32   105   110    32   105   110   100   105    97    46

**Initial key:**
in = GF($2^7$) array. Primitive polynomial = D^7+D^3+1 (137 decimal)

  Array elements =

  Columns 1 through 13

12      4       5       6       7       3       3       5
6       4       4       6       7

  Columns 14 through 26

4       3       5       6       6       7       7       2
4       2       2       5       4

  Columns 27 through 39

6       7       3       3       2       6       7       8
7       8       8       8       9

  Columns 40 through 52

9       4       7       1       7       9       4       5
9       0       2       5       9

  Columns 53 through 63

0       3       5       6       2       6       3       0
1       5       3

**Encryption of first message chunk**:
encry1 = GF($2^7$) array. Primitive polynomial = D^7+D^3+1 (137 decimal)

  Array elements =

  Columns 1 through 13

126     101     111     103     116     119     107
100     104     36      113     104     110

  Columns 14 through 26

114     102     119     117     111     115     126
34      101     108     102     37      119

  Columns 27 through 39

110     117     106     35      111     103     115
105     39      126     105     97      122

  Columns 40 through 52

| 97  | 106 | 104 | 33  | 99  | 108 | 114 |
|-----|-----|-----|-----|-----|-----|-----|
| 108 | 41  | 117 | 108 | 108 | 127 |     |

Columns 53 through 63

| 101 | 113 | 118 | 111 | 118 | 127 | 35 |
|-----|-----|-----|-----|-----|-----|----|
| 97  | 115 | 96  | 35  |     |     |    |

**RS Encoding of first message chunk:**
Code1 = GF(2^7) array. Primitive polynomial = D^7+D^3+1 (137 decimal)

Array elements =

Columns 1 through 13

| 126 | 101 | 111 | 103 | 116 | 119 | 107 |
|-----|-----|-----|-----|-----|-----|-----|
| 100 | 104 | 36  | 113 | 104 | 110 |     |

Columns 14 through 26

| 114 | 102 | 119 | 117 | 111 | 115 | 126 |
|-----|-----|-----|-----|-----|-----|-----|
| 34  | 101 | 108 | 102 | 37  | 119 |     |

Columns 27 through 39

| 110 | 117 | 106 | 35  | 111 | 103 | 115 |
|-----|-----|-----|-----|-----|-----|-----|
| 105 | 39  | 126 | 105 | 97  | 122 |     |

Columns 40 through 52

| 97  | 106 | 104 | 33  | 99  | 108 | 114 |
|-----|-----|-----|-----|-----|-----|-----|
| 108 | 41  | 117 | 108 | 108 | 127 |     |

Columns 53 through 65

| 101 | 113 | 118 | 111 | 118 | 127 | 35 |
|-----|-----|-----|-----|-----|-----|----|
| 97  | 115 | 96  | 35  | 112 | 37  |    |

Columns 66 through 78

| 36  | 92  | 90  | 82  | 123 | 31  | 99  |
|-----|-----|-----|-----|-----|-----|-----|
| 21  | 33  | 41  | 95  | 109 | 101 |     |

Columns 79 through 91

| 100 | 26  | 95  | 19  | 93  | 78  | 86  |
|-----|-----|-----|-----|-----|-----|-----|
| 105 | 18  | 13  | 123 | 91  | 96  |     |

Columns 92 through 104

| 116 | 38  | 43  | 119 | 22  | 11  | 74  |
|-----|-----|-----|-----|-----|-----|-----|
| 46  | 10  | 0   | 54  | 79  | 43  |     |

Columns 105 through 117

| 80  | 48  | 52  | 108 | 2   | 105 | 76  |
|-----|-----|-----|-----|-----|-----|-----|
| 2   | 6   | 21  | 6   | 36  | 102 |     |

Columns 118 through 127

| 8  | 11 | 97 | 24 | 22 | 4 | 58 | 101 |
|----|----|----|----|----|---|----|-----|
| 58 | 65 |    |    |    |   |    |     |

**RS decoding of first message chunk:**
decode1 = GF(2^7) array. Primitive polynomial = D^7+D^3+1 (137 decimal)

Array elements =

Columns 1 through 13

| 126 | 101 | 111 | 103 | 116 | 119 | 107 |
|-----|-----|-----|-----|-----|-----|-----|
| 100 | 104 | 36  | 113 | 104 | 110 |     |

Columns 14 through 26

| 114 | 102 | 119 | 117 | 111 | 115 | 126 |
|-----|-----|-----|-----|-----|-----|-----|
| 34  | 101 | 108 | 102 | 37  | 119 |     |

Columns 27 through 39

| 110 | 117 | 106 | 35  | 111 | 103 | 115 |
|-----|-----|-----|-----|-----|-----|-----|
| 105 | 39  | 126 | 105 | 97  | 122 |     |

Columns 40 through 52

| 97  | 106 | 104 | 33  | 99  | 108 | 114 |
|-----|-----|-----|-----|-----|-----|-----|
| 108 | 41  | 117 | 108 | 108 | 127 |     |

Columns 53 through 63

| 101 | 113 | 118 | 111 | 118 | 127 | 35 |
|-----|-----|-----|-----|-----|-----|----|
| 97  | 115 | 96  | 35  |     |     |    |

**First Message chunk at receiver end:**
msg1 = GF(2^7) array. Primitive polynomial = D^7+D^3+1 (137 decimal)

Array elements =

Columns 1 through 13

| 114 | 97  | 106 | 97  | 115 | 116 | 104 |
|-----|-----|-----|-----|-----|-----|-----|
| 97  | 110 | 32  | 117 | 110 | 105 |     |

Columns 14 through 26

| 118 | 101 | 114 | 115 | 105 | 116 | 121 |
|-----|-----|-----|-----|-----|-----|-----|
| 32  | 97  | 110 | 100 | 32  | 115 |     |

Columns 27 through 39

| 104 | 114 | 105 | 32  | 109 | 97  | 116 |
|-----|-----|-----|-----|-----|-----|-----|
| 97  | 32  | 118 | 97  | 105 | 115 |     |

Columns 40 through 52

| 104 | 110 | 111 | 32  | 100 | 101 | 118 |
|-----|-----|-----|-----|-----|-----|-----|
| 105 | 32  | 117 | 110 | 105 | 118 |     |

Columns 53 through 63

| 101 | 114 | 115 | 105 | 116 | 121 | 32 |
|-----|-----|-----|-----|-----|-----|----|
| 97  | 114 | 101 | 32  |     |     |    |

**New key generated:**
key1 = GF(2^7) array. Primitive polynomial = D^7+D^3+1 (137 decimal)

Array elements =

Columns 1 through 13

| 124 | 33  | 33  | 90  | 93  | 81  | 120 |
|-----|-----|-----|-----|-----|-----|-----|
| 26  | 101 | 17  | 37  | 47  | 88  |     |

Columns 14 through 26

| 105 | 102 | 97  | 28  | 89  | 20  | 90  |
|-----|-----|-----|-----|-----|-----|-----|
| 76  | 82  | 107 | 16  | 8   | 127 |     |

Columns 27 through 39

| 93  | 103 | 119 | 37  | 41  | 113 | 17  |
|-----|-----|-----|-----|-----|-----|-----|
| 3   | 77  | 38  | 2   | 8   | 63  |     |

Columns 40 through 52

| 70  | 47  | 87  | 49  | 51  | 101 | 6   |
|-----|-----|-----|-----|-----|-----|-----|
| 108 | 69  | 2   | 4   | 16  | 15  |     |

Columns 53 through 63

| 36 | 101 | 13 | 13 | 99 | 30 | 21 |
|----|-----|----|----|----|----|----|
| 4  | 59  | 96 | 57 |    |    |    |

**Encryption of second message chunk**:
encry2 = GF(2^7) array. Primitive polynomial = D^7+D^3+1 (137 decimal)

Array elements =

Columns 1 through 13

| 8 | 86 | 78 | 122 | 58 | 62 | 23 |
|---|----|----|-----|----|----|----|
| 126 | 69 | 100 | 75 | 70 | 46 | |

Columns 14 through 26

| 12 | 20 | 18 | 117 | 45 | 109 | 122 |
|----|----|----|-----|----|-----|-----|
| 37 | 60 | 75 | 121 | 102 | 27 | |

Columns 27 through 39

| 52 | 6 | 89 | 37 | 41 | 113 | 17 |
|----|---|----|----|----|-----|----|
| 3 | 77 | 38 | 2 | 8 | 63 | |

Columns 40 through 52

| 70 | 47 | 87 | 49 | 51 | 101 | 6 |
|----|----|----|----|----|-----|---|
| 108 | 69 | 2 | 4 | 16 | 15 | |

Columns 53 through 63

| 36 | 101 | 13 | 13 | 99 | 30 | 21 |
|----|-----|----|----|----|----|----|
| 4 | 59 | 96 | 57 | | | |

**RS encoding of second message chunk**:
Code2 = GF(2^7) array. Primitive polynomial = D^7+D^3+1 (137 decimal)

Array elements =

Columns 1 through 13

| 8 | 86 | 78 | 122 | 58 | 62 | 23 |
|---|----|----|-----|----|----|----|
| 126 | 69 | 100 | 75 | 70 | 46 | |

Columns 14 through 26

| 12 | 20 | 18 | 117 | 45 | 109 | 122 |
|----|----|----|-----|----|-----|-----|
| 37 | 60 | 75 | 121 | 102 | 27 | |

Columns 27 through 39

| 52 | 6 | 89 | 37 | 41 | 113 | 17 |
|----|---|----|----|----|-----|----|
| 3 | 77 | 38 | 2 | 8 | 63 | |

Columns 40 through 52

| 70 | 47 | 87 | 49 | 51 | 101 | 6 |
|----|----|----|----|----|-----|---|
| 108 | 69 | 2 | 4 | 16 | 15 | |

Columns 53 through 65

| 36 | 101 | 13 | 13 | 99 | 30 | 21 |
|----|-----|----|----|----|----|----|
| 4 | 59 | 96 | 57 | 94 | 14 | |

Columns 66 through 78

| 111 | 123 | 43 | 3 | 99 | 76 | 76 |
|-----|-----|----|---|----|----|----|
| 54 | 76 | 37 | 6 | 15 | 104 | |

Columns 79 through 91

| 97 | 23 | 103 | 33 | 0 | 70 | 71 |
|----|----|-----|----|---|----|----|
| 23 | 106 | 26 | 50 | 29 | 85 | |

Columns 92 through 104

| 3 | 71 | 33 | 0 | 95 | 78 | 34 | 125 |
|---|----|----|---|----|----|----|-----|
| 53 | 1 | 32 | 50 | 112 | | | |

Columns 105 through 117

| 9 | 38 | 49 | 67 | 56 | 84 | 23 | 45 |
|---|----|----|----|----|----|----|----|
| 127 | 51 | 51 | 29 | 33 | | | |

Columns 118 through 127

| 47 | 16 | 10 | 10 | 118 | 71 | 79 |
|----|----|----|----|-----|----|----|
| 60 | 13 | 124 | | | | |

**RS decoding of second chunk at receiver end**:
decode2 = GF(2^7) array. Primitive polynomial = D^7+D^3+1 (137 decimal)

Array elements =

Columns 1 through 13

| 8 | 86 | 78 | 122 | 58 | 62 | 23 |
|---|----|----|-----|----|----|----|
| 126 | 69 | 100 | 75 | 70 | 46 | |

Columns 14 through 26

| 12 | 20 | 18 | 117 | 45 | 109 | 122 |
|----|----|----|-----|----|-----|-----|
| 37 | 60 | 75 | 121 | 102 | 27 | |

Columns 27 through 39

| 52 | 6 | 89 | 37 | 41 | 113 | 17 |
|----|---|----|----|----|-----|----|
| 3 | 77 | 38 | 2 | 8 | 63 | |

Columns 40 through 52

| 70 | 47 | 87 | 49 | 51 | 101 | 6 |
|----|----|----|----|----|-----|---|
| 108 | 69 | 2 | 4 | 16 | 15 | |

Columns 53 through 63

| 36 | 101 | 13 | 13 | 99 | 30 | 21 |
|----|-----|----|----|----|----|----|
| 4 | 59 | 96 | 57 | | | |

# 6. CONCLUSION & FUTURE WORK

In this paper, a novel key generation algorithm with the help of forward error correction technique is implemented. In analysis, Reed Solomon code for the generation of dynamic keys is used. While performing this operation, it is noticed that it is very difficult to break generated key because every time there are different key sequences generated. The proposed work provides a solution towards symmetric key cryptography key generation problem. In addition, key refreshment problem is also solved by the suggested techniques with very little efforts. For the simulation purpose in the analysis chunks of block size was taken (127, 63).

In future, researcher can extend and evaluate the proposed work in the following ways:

a) Performance Evaluation with different encryption algorithm

b) Enhancing Robustness through Various Error Control Codes.